\documentclass[reprint,
superscriptaddress,
amsmath,amssymb,
aps, 
pra,
twocolumn
]{revtex4-2}

\usepackage{graphicx}
\usepackage{physics,mathtools}
\usepackage[normalem]{ulem} 

\usepackage{hyperref}
\hypersetup{%
   colorlinks = true,
   linkcolor=blue,
   citecolor=blue,
   urlcolor=blue
 }
 
\renewcommand{\Re}[1]{\text{Re}#1}

\begin{document}
\title{Effective Mass in Dissipative Coupled Polaritons}
\author{D. A. Mendoza} 
\affiliation{Departamento de F\'isica, Universidad Aut\'onoma Metropolitana-Iztapalapa, Av. Ferrocarril San Rafael Atlixco 186, C.P. 09310 Mexico City, Mexico}
\author{A. J. Vega-Carmona} 
\affiliation{Departamento de F\'isica, Universidad Aut\'onoma Metropolitana-Iztapalapa, Av. Ferrocarril San Rafael Atlixco 186, C.P. 09310 Mexico City, Mexico}
\author{A. Camacho-Guardian}
\affiliation{Instituto de F\'isica, Universidad Nacional Aut\'onoma de M\'exico, Ciudad de M\'exico C.P. 04510, Mexico}
\author{Miguel Angel Bastarrachea-Magnani}
\affiliation{Departamento de F\'isica, Universidad Aut\'onoma Metropolitana-Iztapalapa, Av. Ferrocarril San Rafael Atlixco 186, C.P. 09310 Mexico City, Mexico}
\email{bastarrachea@xanum.uam.mx}

\begin{abstract}
Dissipative coupling refers to the effect where two systems interact with each other mediated by dissipation channels. Recent advances in controlling light-matter systems have opened new avenues to explore non-Hermitian effects arising from dissipative coupling, such as level attraction and anomalous dispersions. In this work, we perform a parametric study of these effects in a polariton system, i.e., a light-matter superposition, under both dissipative and coherent coupling. We characterize the effects of different sources of non-Hermitian behavior and analytically identify the conditions for the emergence of negative effective mass, exceptional points, and bound states in the continuum as a function of the light-matter detuning, the coherent-to-dissipative coupling ratio, and the relative decay rate of the non-interacting subsystems. We also analyze the classical limit of the polariton system within a non-Hermitian framework, employing coherent states.
\end{abstract}

\pacs{}

\maketitle

\section{Introduction}
\label{sec:1}

Coherent coupling describes the efficient, non-dissipative energy exchange between quantum degrees of freedom. The control and enhancement of coherent coupling in light-matter quantum systems have been central over the last decades to address fundamental questions, generate novel states of light and matter, and develop fast and versatile applications~\cite{FriskKockum2019,FornDiaz2019,LeBoite2020,Qin2024}. Particularly, these studies have focused on the strong coupling regime, where coherent coupling is larger than the dissipation rates of the systems~\cite{Khitrova2006}. Instead, dissipative coupling, proper of non-Hermitian systems~\cite{Bender2007,Ashida2020}, is associated with an interaction indirectly mediated by dissipation through a common reservoir. In recent years, dissipative coupling has become attractive because it leads to exotic phenomena such as anomalous dispersion relations for quantum particles, level attraction, negative mass, and exceptional points, paving the way for new avenues in tunable interacting quantum systems and quantum technologies~\cite{Harder2021}.  

While coherent coupling is usually understood in terms of level repulsion, as the energy cost for excitations, level attraction or clustering is a feature present in non-Hermitian systems~\cite{Harder2018}, and is connected to resonance trapping~\cite{Persson2000}. It is often accompanied by two other effects. First, anomalous dispersion relations, i.e., where excitation energies change in curvature as a function of momentum, are different from the standard linear or quadratic behaviors, including zeros and curvature inversions~\cite{Colas2018}. This impacts the derivatives of the dispersion, i.e., the effective mass parameters~\cite{Larson2005}, leading, e.g., to a negative mass. Negative mass produces the group velocity to invert its direction, becoming opposite to momentum~\cite{Khamehchi2017}, solitons~\cite{Egorov2009}, and self-interference~\cite{Colas2016}. A second feature is the presence of singular behavior such as exceptional or Diabolic points, a type of singularity where eigenvectors coalesce as a function of the parameter space~\cite{Heiss2012,Tserkovnyak2020}, or bound-states in the continuum, i.e., confined modes embedded in a continuum~\cite{Yang2020,Friedrich1985,Ardizzone2022}. Dissipative coupling allows for potentially controlling the dispersion relation, hence the effective mass and quasiparticle propagation, as well as potential applications such as topological information processing~\cite{Bernier2018,Yang2019}, synchronization~\cite{Grigoryan2018}, non-reciprocal transmission~\cite{Metelmann2015,Wang2019}, and coupling distant systems~\cite{Rao2019,Grigoryan2019}. Likewise, the singularities resulting from these non-Hermitian effects may dramatically change the propagation and topological properties~\cite{Heiss2000,Demboski2004,Zhang2018,Doppler2016,Xu2016}, and enhance sensing~\cite{Liao2021}. 

Dissipative coupling has been observed in several systems such as quantum dots~\cite{Tawara2010,Dalacu2010,Valente2014}, microwave cavities~\cite{Persson2000,Okolowicz2003,Grigoryan2018}, opto-mechano-fluidic resonators~\cite{Lu2023}, and optomechanical systems~\cite{Elste2009,Bernier2018}. Recently, a broad class of tunable light-matter-interacting systems has enabled the manipulation of dissipative coupling. This includes cavity magnonics~\cite{Harder2018,Bhoi2019,Yao2019,Yang2019,Wang2020,Boventer2020,Harder2021} and exciton-polariton systems~\cite{Bieganska2024}. Polaritons are quantum states resulting from the strong light-matter interaction that inherit properties of their original constituents~\cite{Hopfield1958}. Particularly, exciton-polaritons formed in microcavity semiconductors~\cite{Carusotto2013}, have become a versatile tool exploiting their hybrid nature that allows for the exchange of properties between light and matter and create novel and tunable phenomena, including condensates and superfluids~\cite{Amo2009,Deng2010,Kasprzak2006}, topological states~\cite{Ozawa2019}, strongly correlated states~\cite{Takemura2014,Basov2021}, and proposals for qubit implementations~\cite{Ghosh2020,Kavokin2022,Barrat2024}. The versatility, high tunability, and driven-dissipative nature of exciton-polariton systems make them promising for the exploration of non-Hermitian systems~\cite{Bloch2022}. There have been theoretical proposals~\cite{Aleiner2012,Kyriienko2014}, and experimental observations of dissipative coupling effects in trion-polaritons~\cite{Dhara2018}, monolayers of transition metal dichalcogenides (TMDs)~\cite{Wurdack2023}, and AlGaAs-based systems~\cite{Bieganska2024}. Likewise, exceptional points in exciton-polariton systems have been investigated in several works~\cite{Gao2015,Gao2018,Hanai2019,Yu2021,Li2022,Wingenbach2024}.

The origin of dissipative coupling may vary from system to system, and there is no general microscopic framework that explains it. Instead, various approaches have been proposed to address it~\cite{Wang2020}. Classically, it can be explained in terms of classical coupled oscillators as synchronization~\cite{Harder2021} or, in magnon systems, via e.g. the cavity Lenz effect~\cite{Harder2018}. In the quantum limit, a standard proposal is to assume that the dissipative coupling between two degrees of freedom coupled coherently to each other emerges from their interactions with a common environment~\cite{Yu2019}. An input-output approach has been applied to exciton-polariton systems, where the dissipative coupling is explained in terms of the decaying rates of the system~\cite{Bleu2024}.

Inspired by recent developments in polariton systems, in this work, we perform a parametric study of polariton branches combining both coherent and dissipative coupling. We explore the effective mass of polariton branches as a function of the parameter space, namely, the ratio of dissipative to coherent coupling, and the independent photonic and excitonic decay rates. By studying the light-matter detuning and the coherent-to-dissipative coupling ratio, we identify analytic conditions for the emergence of the non-Hermitian features resulting from dissipative coupling. Moreover, we analyze the semi-classical corresponding system to relate the findings at the mean-field level. 

The manuscript is organized as follows. In Section~\ref{sec:2}, we introduce the system, the polariton basis, and discuss the concepts of inertial and diffusive masses. In Section~\ref{sec:3}, we study the behavior of the polariton branches as a function of decay and the dissipative coupling, as well as their combined effect, characterizing the parameter space via the effective masses. Also, we study conditions for the onset of exceptional points (EPs) and bound states in the continuum. In Section~\ref {sec:4}, we study the classical limit via coherent states. Finally, in Section~\ref{sec:5} we present our conclusions and offer some perspectives.


\section{Non-Hermitian system}
\label{sec:2}

We consider exciton-polaritons formed in a microcavity semiconductor, where excitons strongly couple to cavity photons, and both experience dissipation mediated by the surrounding environment. The Hamiltonian reads
\begin{gather}
    \hat{H}=\sum_{\textbf{k}}\begin{bmatrix}
        \hat{c}_{\textbf{k}}^{\dagger} & \hat{x}_{\textbf{k}}^{\dagger}
        \end{bmatrix}\begin{bmatrix}
        \varepsilon_{c\textbf{k}}-i\gamma_{c} & \Omega-i\Omega_{\text{Im}}\\
        \Omega-i\Omega_{\text{Im}} & \varepsilon_{x\textbf{k}}-i\gamma_{x}
        \end{bmatrix}\begin{bmatrix}
          \hat{c}_{\textbf{k}}\\
          \hat{x}_{\textbf{k}}
        \end{bmatrix},
\end{gather}
where $\hat{x}_{\mathbf{k}}^{\dagger}$ and $\hat{c}_{\mathbf{k}}^{\dagger}$ are the exciton and photon annihilation operators for momentum $\mathbf{k}$, and dispersion relations $\varepsilon_{x\mathbf{k}}=\mathbf{k}^{2}/2m_{x}$ and $\varepsilon_{c\mathbf{k}}=\mathbf{k}^{2}/2m_{c}+\delta$, being $m_{x}$ and $m_{c}$ their masses and  $\delta$ is the zero-momentum light-matter detuning. We take $m_{c}=10^{-4}m_{x}$, $m_{x}=m_{e}/2$, where $m_{e}$ is the electron mass. $\Omega$ is the light-matter coherent coupling. We incorporate dissipation in the system by adding non-Hermitian terms to the Hamiltonian, accounting for the decay of uncoupled excitons and photons, characterized by the parameters $\gamma_x$ and $\gamma_c$, respectively, and the dissipative coupling represented by the parameter $\Omega_{\text{Im}}$.

\subsection{Dissipative coupling polariton branches}

We diagonalize the Hamiltonian by means of a Hopfield transformation~\cite{Hopfield1958}
\begin{align}
\begin{bmatrix}
\hat{c}_{\mathbf{k}} \\
\hat{x}_{\mathbf{k}}
\end{bmatrix}
=
\begin{bmatrix}
\tilde{\mathcal{C}}_{\mathbf{k}} & -\tilde{\mathcal{S}}_{\mathbf{k}} \\
\tilde{\mathcal{S}}_{\mathbf{k}} &  \tilde{\mathcal{C}}_{\mathbf{k}}
\end{bmatrix}
\begin{bmatrix}
\hat{U}_{\mathbf{k}} \\
\hat{L}_{\mathbf{k}}
\end{bmatrix},
\label{eq:Polaritons}
\end{align}
where $L_{\mathbf{k}}^{\dagger}$ ($U_{\mathbf{k}}^{\dagger}$) are the creation operators of the dissipative coupling lower and upper polaritons, respectively, with momentum $\mathbf{k}$. The presence of dissipation modifies the Hopfield coefficients in the following way
\begin{gather} \label{eq:hop1}
    \tilde{\mathcal{C}}_{\textbf{k}}^{2}=\frac{1}{2}\left(1+\frac{\tilde\delta_{\textbf{k}}}{\sqrt{\tilde\delta_{\textbf{k}}^{2}+4(\Omega-i\Omega_{\text{Im}})^{2}}}\right), \\ \nonumber
    \tilde{\mathcal{S}}_{\textbf{k}}^{2}=1-\tilde{\mathcal{C}}_{\textbf{k}}^{2},
\end{gather}
where $\tilde{\delta}_{\textbf{k}}=\left(\varepsilon_{c\mathbf{k}}-i\gamma_c\right)-\left(\varepsilon_{x\mathbf{k}}-i\gamma_x\right)=\delta_{\textbf{k}}-i\Delta\gamma$, being $\delta_{\mathbf{k}}=\varepsilon_{c\mathbf{k}}-\varepsilon_{x\mathbf{k}}$ the finite-momentum light-matter detuning and $\Delta\gamma=\gamma_c-\gamma_x$ is the relative decay rate. Polariton branches are now identified by the complex eigenvalues of the Hamiltonian, given by
\begin{gather} \label{eq:hop2}
    E_{\sigma\textbf{k}}=\frac{1}{2}\left(\delta_{\textbf{k}}+2\varepsilon_{x\textbf{k}}-i(\gamma_{x}+\gamma_{c})\pm\sqrt{\tilde{\delta}_{\mathbf{k}}^{2}+4(\Omega-i\Omega_{\text{im}})^{2}}\right),
\end{gather}
where $\sigma=\text{L},\text{U}$ denotes the lower ($\text{L}$) and upper ($\text{U}$) exciton-polaritons branches. We observe that, from Eqs.~\ref{eq:hop1} and~\ref{eq:hop2}, by making all the sources of dissipation equal to zero one recovers the standard Hopfield coefficients $\mathcal{C}_{\mathbf{k}}^{2}=(1+\delta_{\mathbf{k}}/\sqrt{\delta_{\mathbf{k}}^{2}+4\Omega^{2}})/2$ and $\mathcal{S}_{\mathbf{k}}^{2}=1-\mathcal{C}_{\mathbf{k}}^{2}$, and polariton energies $E_{\sigma\mathbf{k}}=\frac12\left(\delta_{\mathbf{k}}+2\varepsilon_{x\mathbf{k}}\pm\sqrt{\delta_{\mathbf{k}}^{2}+4\Omega^{2}}\right)$. 

The physics of the dissipative coupling are encoded in the polariton branches~\cite{Harder2017}, whose energy can be analytically separated into its real and imaginary parts, without making any approximations, as $E_{\sigma\mathbf{k}}=\varepsilon_{\sigma\mathbf{k}}-i\gamma_{\sigma\mathbf{k}}$, with
\begin{gather}
    \varepsilon_{\sigma\textbf{k}}=\frac{1}{2}\left(2\varepsilon_{x\textbf{k}} +\delta_{\textbf{k}}\pm \mathcal{F}_{\mathbf{k}}^{+}(\Omega_{\text{Im}},\Delta\gamma,\delta)\right),
\end{gather}
and
\begin{gather}
    \gamma_{\sigma\textbf{k}}=\frac{1}{2}\left(\gamma_{x}+\gamma_{c}\right)\\ \nonumber
    \pm\frac{1}{4}\left[\text{sign}\left[\mathcal{J}_{\mathbf{k}}(\Omega_{\text{Im}},\Delta\gamma,\delta)\right]\mathcal{F}_{\mathbf{k}}^{-}(\Omega_{\text{Im}},\Delta\gamma,\delta)\right]
\end{gather}
where
\begin{gather}
\mathcal{F}_{\mathbf{k}}^{\pm}(\Omega_{\text{Im}},\Delta\gamma,\delta)=\frac{1}{\sqrt{2}}\sqrt{\mathcal{G}_{\mathbf{k}}\pm \mathcal{H}_{\mathbf{k}}}, \\
\mathcal{G}_{\mathbf{k}}(\Omega_{\text{Im}},\Delta\gamma,\delta)=\sqrt{\mathcal{H}_{\mathbf{k}}^{2}+\mathcal{J}_{\mathbf{k}}^{2}},\\
\mathcal{H}_{\mathbf{k}}(\Omega_{\text{Im}},\Delta\gamma,\delta)=\delta_{\textbf{k}}^{2}-\Delta\gamma^{2}+4(\Omega^{2}-\Omega_{\text{Im}}^{2}),\\
\mathcal{J}(\Omega_{\text{Im}},\Delta\gamma,\delta)=2\left(\delta_{\textbf{k}}\Delta\gamma+4\Omega\Omega_{\text{Im}}\right).
\end{gather}

Using the above result, we can express the diagonalized Hamiltonian as $\hat{H}=\hat{H}_{\text{R}}+i\hat{K}$, and 
\begin{gather}
\hat{H}_{\text{R}}=\sum_{\mathbf{k}}\hat{H}_{\mathbf{k}}=\sum_{\mathbf{k}}\epsilon_{\text{LP}\mathbf{k}}\hat{L}_{\mathbf{k}}^{\dagger}\hat{L}_{\mathbf{k}}+\epsilon_{\text{UP}\mathbf{k}}\hat{U}_{\mathbf{k}}^{\dagger}\hat{U}_{\mathbf{k}}, \label{eq:1}\\
\hat{K}=\sum_{\mathbf{k}}\hat{K}_{\mathbf{k}}=\sum_{\mathbf{k}}\gamma_{\text{LP}\mathbf{k}}\hat{L}_{\mathbf{k}}^{\dagger}\hat{L}_{\mathbf{k}}+\gamma_{\text{UP}\mathbf{k}}\hat{U}_{\mathbf{k}}^{\dagger}\hat{U}_{\mathbf{k}}, \label{eq:2}
\end{gather}
where $\hat{K}$ accounts for the non-Hermitian part of the Hamiltonian.

Often, one can approximate the above expressions by assuming that dissipation is smaller than the coherent energy exchange. This occurs, for example, when one considers that the dissipation in the photonic part arises from imperfections in the mirrors inside the cavity, so it can be well-characterized, being $\gamma_{c}/2\Omega\ll1$. A similar situation can be supposed for the excitonic dissipation. In such a case, we can approximate the energy expressions to obtain the following dissipative coupling polariton branches
\begin{gather}\label{eq:enrgdisp}
       \varepsilon_{\sigma\textbf{k}}- i\gamma_{\sigma\textbf{k}}\approx\frac{1}{2}\delta_{\textbf{k}}+\varepsilon_{x\textbf{k}}-\frac{i}{2}(\gamma_{x}+\gamma_{c})\\ \nonumber
       \pm\frac{1}{2}\sqrt{\delta^2_{\textbf{k}}+4(\Omega-i\Omega_{\text{Im}})^{2}} -\frac{i}{2}\frac{\delta_{\textbf{k}}(\gamma_{c}-\gamma_{x})}{\sqrt{\delta^2_{\textbf{k}}+4(\Omega-i\Omega_{\text{Im}})^{2}}}\\ \nonumber
       =\frac{1}{2}\left(\delta_{\textbf{k}}+2\varepsilon_{\textbf{k}}^{x}\pm\sqrt{\delta_{\textbf{k}}^2+4(\Omega-i\Omega_{\text{Im}})^{2}}\right)\\ \nonumber
       -i\left[\frac{\gamma_{x}}{2}\left(1\pm\frac{\delta_{\textbf{k}}}{\sqrt{\delta_{\textbf{k}}^{2}+4(\Omega-i\Omega_{\text{Im}})^{2}}}\right)\right.+\\ \nonumber
       \left.\frac{\gamma_{c}}{2}\left(1\mp\frac{\delta_{\textbf{k}}}{\sqrt{\delta_{\textbf{k}}^{2}+4(\Omega-i\Omega_{\text{Im}})^{2}}}\right)\right]
    \end{gather}
Therefore one can define a dissipation rate for each polariton branch as $\gamma_{\textbf{k}}^{\text{UP}}=\gamma_{x}\tilde{C}_{\textbf{k}}^{2}+\gamma_{c}\tilde{S}_{\textbf{k}}^{2}$ and $\gamma_{\textbf{k}}^{\text{LP}}=\gamma_{x}\tilde{S}_{\textbf{k}}^{2}+\gamma_{c}\tilde{C}_{\textbf{k}}^{2}$.

\subsection{Inertial and diffusion masses}

In solid-state physics, the classical concept of inertial mass is insufficient to describe the dynamics of particles in a medium where interactions and collective effects play a significant role. The concept of \textit{effective mass} is introduced to capture the modified inertial response of quasiparticles within such environments. This effective mass is defined in terms of the system’s energy as a function of momentum. By performing a Taylor expansion of the energy–momentum relation around a given point, the effective mass can be related to the first and second derivatives of the dispersion relation, with the second derivative being particularly relevant for determining how the particle accelerates under external forces~\cite{Larson2005,Colas2018}. It reads, 
\begin{gather}
   \varepsilon_{\sigma\textbf{k}} \approx \varepsilon_{\sigma 0}
+ \mathbf{k}\cdot \left.\nabla_{\mathbf{k}} \varepsilon_{\sigma\textbf{k}}\right|_{\mathbf{k}=0}
+ \frac{1}{2}\,\mathbf{k}\cdot\left.\mathbf{H}(\varepsilon_{\sigma\textbf{k}})\right.|_{\mathbf{k}=\mathbf{0}}
\cdot\mathbf{k}
+ \dots
\end{gather}
where $\mathbf{H}(\varepsilon_{\sigma\textbf{k}})$ is the Hessian matrix, whose elements are $[\mathbf{H}(\varepsilon_{\sigma\mathbf{k}})]_{ij}=\partial^{2}\varepsilon_{\sigma\mathbf{k}}/\partial k_{i}\partial k_{j}$ with $i,j=x,y,z$. The coefficients of each order term are related to a new mass parameter that has certain characteristic effects on the dynamics. For an isotropic medium, we define the mass parameters as
\begin{gather}\label{eq:mm}
m_{1}^{-1} = \frac{\mathbf{k}}{||\mathbf{k}||}\,
\cdot \left.\nabla_{\mathbf{k}}\varepsilon_{\sigma\textbf{k}}\right|_{\mathbf{k}=0},
\\ \nonumber
m_{2}^{-1} \equiv (m^{*})^{-1}
= \left.\nabla_{\mathbf{k}}^{2}\varepsilon_{\sigma\textbf{k}}\right|_{\mathbf{k}=\mathbf{0}}.
\end{gather}
The inertial mass $m_{1}$ parameter is related to the group velocity $v_{g}=|\mathbf{k}|/m_{1}$. Instead, the diffusive mass $m_{2}$ parameter determines the particle's acceleration under an external force, and the rate of diffusion, hence the name~\cite{Colas2018}.

Here, we are primarily interested in the diffusive mass, because variations in the second derivative of the dispersion relation customarily correspond to changes in the value of the effective mass $m^{*}$. As it depends on the light-matter detuning $\delta$, it can be experimentally controlled. We compute the momentum derivatives of the full complex dispersion $E_{\sigma\mathbf{k}}$, and later take their real parts when extracting 
the effective mass, since only the curvature of $\varepsilon_{\sigma\mathbf{k}} 
= \Re[E_{\sigma\mathbf{k}}]$ enters the definition of $m^{*}$.
\begin{gather}\label{eq:d1}    \mathbf{k}\cdot\nabla_{\mathbf{k}}E_{\sigma\mathbf{k}}=\frac{1}{2}\left( \frac{\textbf{k}^{2}}{m_{c}}+\frac{\textbf{k}^{2}}{m_{x}}\right)\\ \nonumber
\pm \frac{1}{2}\tilde{\delta}_{\textbf{k}}\left(\tilde{\delta}_{\textbf{k}}^{2}+4(\Omega-i\Omega_{\text{Im}})^{2}\right)^{-1/2} \left(\frac{\textbf{k}^{2}}{m_{c}}-\frac{\textbf{k}^{2}}{m_{x}}\right), 
\end{gather}
whose real and imaginary parts are given in App.~\ref{app:1}. Meanwhile, the second derivative is
\begin{gather}  \label{eq:m2}
\nabla^{2}_{\mathbf{k}}E_{\sigma\mathbf{k}}=\frac{1}{2}\left(\frac{1}{m_{c}}+\frac{1}{m_{x}} \right)\\ \nonumber
\pm\frac{1}{2}
    \left(\tilde{\delta}_{\textbf{k}}^{2}+4(\Omega-i\Omega_{\text{Im}})^{2}\right)^{-1/2}\left|\frac{\textbf{k}}{m_{c}}-\frac{\textbf{k}}{ m_{x}}\right|^{2}\\ \nonumber
    \pm\frac{1}{2}\tilde{\delta}_{\textbf{k}}\left(\tilde{\delta}_{\textbf{k}}^{2}+4(\Omega-i\Omega_{\text{Im}})^{2}\right)^{-1/2}\left(\frac{1}{ m_{c}}-\frac{1}{ m_{x}}\right)\\ \nonumber
    \mp\frac{1}{2}\tilde{\delta}_{\textbf{k}}^{2}\left(\tilde{\delta}_{\textbf{k}}^{2}+4(\Omega-i\Omega_{\text{Im}})^{2}\right)^{-3/2}\left|\frac{\textbf{k}}{ m_{c}}-\frac{\textbf{k}}{ m_{x}}\right|^{2}
\end{gather}
whose real and imaginary parts are given in App.~\ref{app:1} as well. Because we are interested in the effective mass, we evaluate equation Eq.~\ref{eq:mm} in $\mathbf{k}=\mathbf{0}$:

\begin{gather}
\nabla^{2}_{\mathbf{k}}E_{\sigma\textbf{0}}
        =\frac{1}{2}\left(\frac{1}{m_{c}}+\frac{1}{m_{x}} \right)\\ \nonumber
        \pm\frac{1}{2}\left(\frac{1}{ m_{c}}-\frac{1}{ m_{x}}\right)\frac{\delta-i\Delta\gamma}{\sqrt{\left(\delta-i\Delta\gamma\right)^{2}+4(\Omega-i\Omega_{\text{Im}})^{2}}}.
\end{gather}

We identify the Hopfield coefficients in this last equation. Therefore, the effective masses for the UP and the LP are given by
\begin{gather}
     m_{\text{UP}}^{*}=\left(\frac{\tilde{\mathcal{C}}^{2}}{m_{c}}+\frac{\tilde{\mathcal{S}}^{2}}{m_x}\right)^{-1}, \\ \nonumber
     m_{\text{LP}}^{*}=\left(\frac{\tilde{\mathcal{S}}^{2}}{m_{c}}+\frac{\tilde{\mathcal{C}}^{2}}{m_x}\right)^{-1}.
\end{gather}
Here, $\tilde{\mathcal{C}}^{2}$ and $\tilde{\mathcal{S}}^{2}$ denote the real parts of the  dissipative Hopfield coefficients evaluated at $\mathbf{k}=\mathbf{0}$, i.e. $\tilde{C}^{2} = \Re[\tilde{\mathcal{C}}_{\mathbf{0}}^{2}]$ and $\tilde{S}^{2} = \Re[\tilde{\mathcal{S}}_{\mathbf{0}}^{2}]$, since  the effective mass is defined from the curvature of the real part of the dispersion.

\section{Non-hermitian approach}
\label{sec:3}

To grasp how each source of dissipation modifies the behavior of the system, in this section, we will analyze the parametric behavior of the polariton branches and the effective mass in three cases
\begin{itemize}
    \item \textbf{Case 1:} There is dissipation only in the non-interacting systems $\gamma_{x},\gamma_{c}\neq0$, the coupling is coherent $\Omega_{\text{Im}}=0$.
    \item \textbf{Case 2:} There is only dissipative coupling $\Omega_{\text{Im}}\neq0$, the individual dissipation in the non-interacting systems is zero, $\gamma_{x},\gamma_{c}=0$.
    \item \textbf{Case 3:} There is dissipation in both the non-interacting systems $\gamma_{x},\gamma_{c}\neq0$ and dissipative coupling $\Omega_{\text{Im}}\neq0$, so we have a combined case.
\end{itemize}

\begin{figure*}[!ht]
\begin{center}
\begin{tabular}{c}
\includegraphics[width=0.7\textwidth]{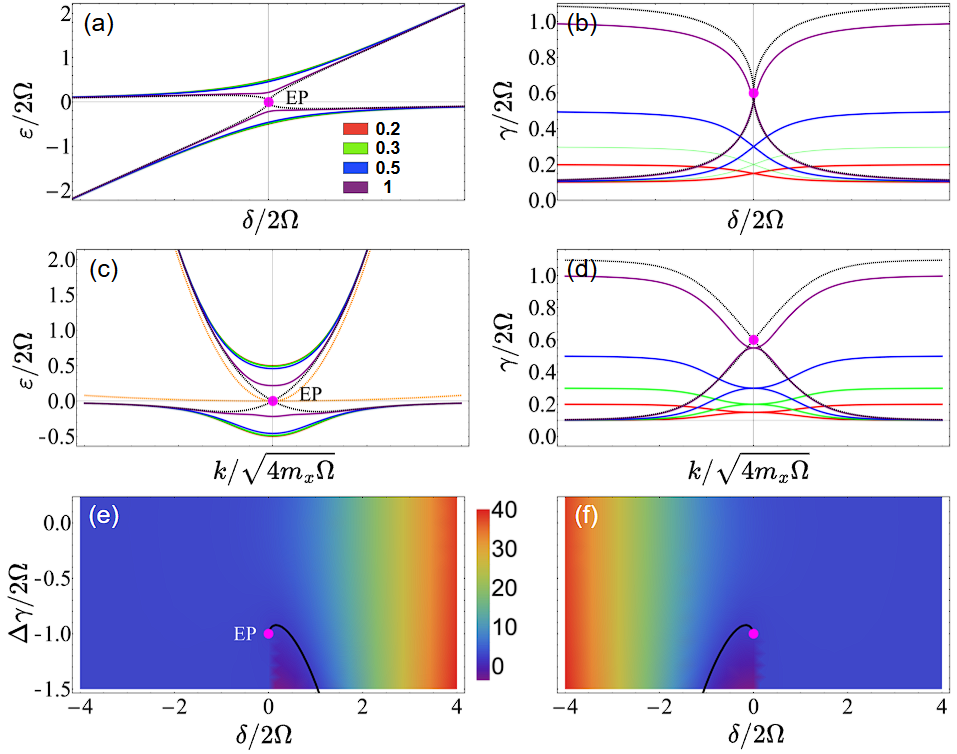}
\end{tabular}
\end{center}
\vspace{-20pt}
\caption{(a) Real and (b) imaginary parts of the exciton-polariton energy as a function of detuning, and (c) and (d) as a function of momentum. In panels (a)–(d), different colors represent curves corresponding to different values of $\gamma_{x}/2\Omega=0.2$, $0.3$, $0.5$, $1.0$, and where the EPs occur. The orange color corresponds to the bare exciton and photon. In the third row, we show the density map of the effective mass $m^{*}$ as a function of detuning $\delta$ and the relative dissipation $\Delta\gamma$ for the (e) upper and (f) lower polariton branches. In panels (e) and (f), the solid black lines indicate the points where the effective mass vanishes. Throughout all panels, we have fixed $\gamma_{c}/2\Omega=0.1$.}
\label{fig:1} 
\end{figure*}

\subsection{Effect of dissipation in the non-interacting systems}

First, we revisit the dissipative effect on the polariton branches when we consider only the dissipation in the non-interacting systems. That is, the photonic component may experience losses due to imperfections in the cavity mirrors. Similarly, the excitonic component has a time decay that depends on the specific material involved, vibrations, and thermal phenomena. Hence, here the terms $\gamma_{x}$ and $\gamma_{c}$ are non zero, and the term $\Omega_{\text{Im}}$ is zero. This case has been studied before in several previous works~\cite{Gao2015,Li2022,Wingenbach2024}. 

The analytical expressions of the energies, derived from the expression Eq.~\ref{eq:hop2}, become
\begin{gather}
E_{\sigma\mathbf{k}}=\varepsilon_{\sigma\textbf{k}}- i\gamma_{\sigma\textbf{k}}=\\ \nonumber
\frac{1}{2}\left(\delta_{\textbf{k}}+2\varepsilon_{x\textbf{k}}-i(\gamma_{x}+\gamma_{c})\pm\sqrt{\tilde{\delta}_{\textbf{k}}^{2}+4\Omega^{2}}\right).
\end{gather}
We separate the real and imaginary parts
\begin{gather}
    \varepsilon_{\sigma\textbf{k}}=\frac{1}{2}\left(2\varepsilon_{x\textbf{k}} +\delta_{\textbf{k}}\pm \mathcal{F}_{\mathbf{k}}^{+}(0,\Delta\gamma,\delta)\right),
\end{gather}
\begin{gather}
    \gamma_{\sigma\textbf{k}}=\frac{1}{2}\left(\gamma_{x}+\gamma_{c}\right)\\ \nonumber
    \pm\frac{1}{4}\text{sign}\left[\mathcal{J}_{\mathbf{k}}(0,\Delta\gamma,\delta)\right]\mathcal{F}_{\mathbf{k}}^{-}(0,\Delta\gamma,\delta),
\end{gather}
with
\begin{gather}
\mathcal{F}_{\mathbf{k}}^{\pm}(0,\Delta\gamma,\delta)=\frac{1}{\sqrt{2}}\sqrt{\mathcal{G}_{\mathbf{k}}(0,\Delta\gamma,\delta)\pm\mathcal{H}_{\mathbf{k}}(0,\Delta\gamma,\delta)}, \\
\mathcal{G}_{\mathbf{k}}(0,\Delta\gamma,\delta)=\sqrt{\mathcal{H}_{\mathbf{k}}^{2}(0,\Delta\gamma,\delta)+\mathcal{J}_{\mathbf{k}}^{2}(0,\Delta\gamma,\delta)},\\
\mathcal{H}_{\mathbf{k}}(0,\Delta\gamma,\delta)=\delta_{\textbf{k}}^{2}-\Delta\gamma^{2}+4\Omega^{2},\\
\mathcal{J}_{\mathbf{k}}(0,\Delta\gamma,\delta)=2\delta_{\mathbf{k}}\Delta\gamma.
\end{gather}

We plot these expressions in Fig.~\ref{fig:1}. To this end, we fix the cavity photon decay rate $\gamma_{c}/2\Omega=0.1$, and vary the exciton decay rate $\gamma_{x}/2\Omega$ as a parameter. In Figs.~\ref{fig:1} (a) and (b), we observe the real and imaginary parts of the polariton branches as a function of the zero-momentum detuning $\delta$. The main effect of increasing the decay $\gamma_{x}$ is that the polariton branches attract to each other, decreasing the gap until it vanishes, where one gets EPs. On the other hand, the imaginary part of each branch intercepts the other for the values where the real part of the energy has the maximum approach, i.e., at a zero detuning in this case. The dashed curves for both real and imaginary curves correspond to the EPs. They occur when the two branches become degenerate, i.e., at $E_{\text{LP}\mathbf{k}}=E_{\text{UP}\mathbf{k}}$, leading to the conditions:
\begin{gather}
\delta_{\mathbf{k}}^{\text{EP}}=0,\,\,\,\, 
\Delta{\gamma}^{\text{EP}}=\pm 2\Omega,
\end{gather}
 i.e., the coherent frequencies must be equal, and the relative dissipation equals the light-matter coupling. 
 
In Figs.~\ref{fig:1} (c) and (d), we observe the real and imaginary parts of the dispersion relation for each branch. In this case, there is no deviation from the standard quadratic behavior. This means that dissipation alone does not change the effective mass. However, again, we have EPs, occurring when the above conditions are met. Notice that there is only a single EP as a function of both detuning and momentum. We can say that the effect of decay is to decrease the polariton gap, a natural result given that it diminishes the effect of the light-matter coupling. 
 
Now, in Figs.~\ref{fig:1} (e) and (f), we have plotted a density map of the effective masses for the lower and upper polaritons, respectively, as a function of detuning and relative dissipation $\Delta\gamma$. According to what was observed before, increasing the detuning makes the effective mass of the upper polariton smaller and larger than that of the lower polariton. This results from the well-known change in the light-matter content, as increasing the light content makes the upper polariton lighter, and the lower polariton heavier. However, as shown in Figs.~\ref{fig:1} (e) and (f), increasing relative dissipation beyond $\Delta\gamma=2\Omega$ leads to a condition where the mass becomes zero ($ \Re[m_{\sigma}^{*}]=0$). This condition is calculated numerically and is indicated as a solid black curve in Figs.~\ref{fig:1} (e) and (f). We also notice that, as a function of the light-matter detuning, the zero-mass condition is bounded by the EP, depicted as a pink dot. 

Next, we observe that dissipation in the system favors the appearance of negative mass values in the polariton dispersion: the greater the dissipation, the larger the area where this phenomenon appears. It occurs in regimes where dissipation exceeds coherent coupling. Notice that the dissipative regions where the lower and upper polaritons acquire negative mass are the same for dissipation, but change depending on whether the detuning is negative or positive. In any case, the negative mass is small, given that the changes in curvature of the dispersion relation are not significant.

We also look for bound states in the continuum (BIC), by making the imaginary part of the polariton branches equal to zero. This leads to the expression
\begin{gather}
2\left(\gamma_x+\gamma_c\right)^2=\sqrt{\left[\delta_{\textbf{k}}^{2}-\Delta\gamma^{2}+4\Omega^{2}\right]^{2}+2\delta^2_{\textbf{k}}\Delta\gamma^{2}} \\ \nonumber
-(\delta_{\textbf{k}}^{2}-\Delta\gamma^{2}+4\Omega^{2}),
\end{gather} 
which in turn becomes the following condition in the detuning vs. relative dissipation parameter space:
\begin{gather} 
    \delta_{\textbf{k}}^2+(2\gamma_{c}-\Delta\gamma)^{2}\left(1+\frac{\Omega^{2}}{\gamma_{c}(\gamma_c+\Delta\gamma)}\right)=0.
\end{gather}
This last equation has no real solutions, due to the squared term and the fact that the dissipation parameters are both taken positive $\gamma_{x},\gamma_{c}>0$. So, without dissipative coupling, there is no existence of BICs. As we will see later, they are understood as a balance between dissipative coupling, coherent coupling, and dissipation in non-interacting systems. Therefore, BICs will only appear when all three effects are present. Moreover, we observe that, in this case, the imaginary part of the polariton branches is always positive, so the polaritons are dissipative. 

\subsection{Effect of dissipative coupling}

Now, we evaluate the effect of dissipative coupling alone $\Omega_{\text{Im}}\neq0$, while dissipation in the non-interacting systems remains zero $\gamma_x=\gamma_c=0$. The eigenenergies are:
\begin{gather}
    E_{\sigma\textbf{k}}=\frac{1}{2}\left(\delta_{\textbf{k}}+2\varepsilon_{x\textbf{k}}\pm\sqrt{\delta_{\textbf{k}}^{2}+4(\Omega-i\Omega_{\text{Im}})^{2}}\right).
\end{gather}
Once again, we separate real and imaginary parts:
\begin{gather} \label{eq:poldiss}
    \varepsilon_{\sigma\textbf{k}}=\frac{1}{2}\left(2\varepsilon_{x\textbf{k}} + \delta_{\textbf{k}}\pm \mathcal{F}_{\mathbf{k}}^{-}(\Omega_{\text{Im}},0,\delta)\right),
\end{gather}
\begin{gather}\label{eq:gam2}
    \gamma_{\sigma\textbf{k}}=\pm\frac{1}{2}\mathcal{F}_{\mathbf{k}}^{-}(\Omega_{\text{Im}},0,\delta),
\end{gather}
where
\begin{gather}
\mathcal{F}_{\mathbf{k}}^{\pm}(\Omega_{\text{Im}},0,\delta)=\frac{1}{\sqrt{2}}\sqrt{\mathcal{G}_{\mathbf{k}}^{\pm}(\Omega_{\text{Im}},0,\delta)
\pm\mathcal{H}_{\mathbf{k}}(\Omega_{\text{Im}},0,\delta)}, \\ \nonumber
\mathcal{G}_{\mathbf{k}}^{\pm} (\Omega_{\text{Im}},0,\delta)=\sqrt{\mathcal{H}_{\mathbf{k}}^{2}(\Omega_{\text{Im}},0,\delta)+\mathcal{J}_{\mathbf{k}}^{2}(\Omega_{\text{Im}},0,\delta)},\\
\mathcal{H}_{\mathbf{k}}(\Omega_{\text{Im}},0,\delta)=\delta_{\textbf{k}}^{2}+4(\Omega^{2}-\Omega_{\text{Im}}^{2}),\\
\mathcal{J}_{\mathbf{k}}(\Omega_{\text{Im}},0,\delta)=8\Omega\Omega_{\text{Im}}.
\end{gather}

\begin{figure*}[!ht]
\begin{center}
\begin{tabular}{c}
\includegraphics[width=0.7\textwidth]{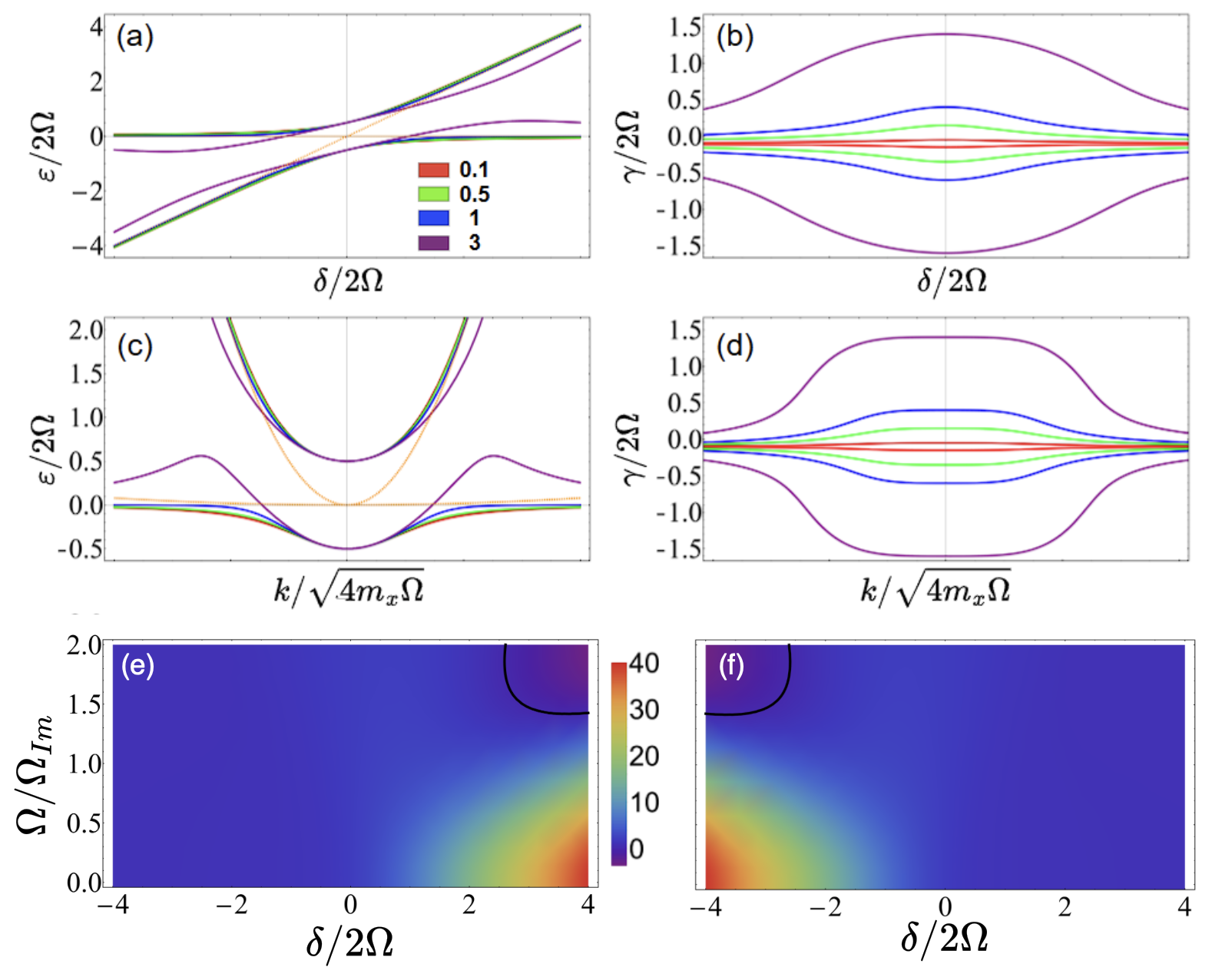}
\end{tabular}
\end{center}
\vspace{-20pt}
\caption{In panels (a)-(d) we show the same as Fig.~\ref{fig:1} (a)-(d), but for different values of the $\tilde{\Omega}=\Omega_{\text{Im}}/\Omega=0.1$, $0.5$, $1.0$, and $3.0$. In the third row, we show the density map of the effective mass $m^{*}$ as a function of detuning $\delta$ and the dissipative coupling $\Omega_{\text{Im}}$ for the (e) upper and (f) lower polariton branches. In panels (e) and (f), the solid black lines indicate the points where the effective mass vanishes. Throughout all panels, we have fixed $\gamma_{c}/2\Omega=0.1$.
}
\label{fig:2} 
\end{figure*}

Like in the previous case, we plot these results in Fig.~\ref{fig:2} as a function of momentum and detuning. In Fig.~\ref{fig:2} (a), we observe that the Rabi splitting does not decrease at zero detuning, so the polariton branches do not cross. Here, the discriminant of the square root in Eq.~\ref{eq:poldiss} is $\delta_{\mathbf{k}}^{2}+4(\Omega^{2}-\Omega_{\mathrm{Im}}^{2})-i\,8\Omega\Omega_{\mathrm{Im}}$, and the existence of EPs require it to vanish. This implies simultaneously $\delta_{\mathbf{k}}^{2}+4(\Omega^{2}-\Omega_{\mathrm{Im}}^{2})=0$ and $\Omega\Omega_{\mathrm{Im}}=0$. Since the light-matter coupling $\Omega\neq 0$, these conditions cannot be satisfied, and therefore dissipative coupling alone does \emph{not} generate EPs. The apparent rapprochement of the real parts in Fig.~\ref{fig:2} (a) reflects a deformation of the dispersion induced by $\Omega_{\mathrm{Im}}$, but it does not correspond to an exceptional point. Setting $\Omega=0$ gives the condition $\delta=\pm\Omega_{\mathrm{Im}}$, but in this limit, the system no longer supports polaritons, and the degeneracy has no physical meaning. We therefore
conclude that dissipative coupling alone does not generate EPs.

In Fig.~\ref{fig:2} (c), we observe that the dissipative coupling produces a noticeable change in the curvature of the dispersion relation, leading to conditions of zero and negative effective mass. However, this only occurs for large momentum. Likewise, the imaginary part in Fig.~\ref{fig:2} (d) exhibits a behavior similar to that as a function of the detuning. 

Again, we explore a density map of the parameter space in Fig.~\ref{fig:2} (e) and (f) for the upper and lower polaritons, respectively. As in the first case, the presence of negative values of effective mass emerges as the dissipative coupling also increases. The black line in the figure represents the points where the effective mass is zero. From Fig.~\ref{fig:2} (c), we observe that negative mass appears first for finite momentum and gets closer to the zero momentum domain with the increase of the dissipative coupling. Hence, negative mass only appears for values where $\Omega_{\text{Im}}\simeq 1.5\Omega$, at positive (negative) detuning for the upper (lower) polariton. 

Finally, we look for BIC under these conditions. By setting the imaginary part of the eigenvalues equal to zero, we get
\begin{gather}
\sqrt{\left[\delta_{\textbf{k}}^{2}+4(\Omega^{2}-\Omega_{\text{Im}}^{2})\right]^{2}+64\Omega^{2}\Omega_{\text{Im}}^{2}}\\ \nonumber
-(\delta_{\textbf{k}}^{2}+4(\Omega^{2}-\Omega_{\text{Im}}^{2}))=0,
\end{gather} 
we observe this expression leads to $\Omega\Omega_{\text{Im}}=0$, which cannot be hold in our current case. So, we cannot obtain BIC from the dissipative coupling parameter alone. However, from Fig.~\ref{fig:2} (b) and (c), we observe that the imaginary part of the lower polariton becomes negative. This means the system is no longer stable, as will be evident later in the semiclassical description. This is explained because in the absence of decay channels for the non-interacting systems, the dissipative coupling serves as an energy pump to the system.

\subsection{Combined effect}

For this last case, we combine the dissipative coupling and the decays of the non-interacting systems. From the results of the previous cases, we know that the decay parameters are responsible for creating level attraction (reduction in the gap between the LP and UP branches) at maximal hybridization ($\delta=0$), so one can also get EPs. Dissipative coupling is a parameter that creates level attraction in the zero-detuning neighborhood, as well as negative mass at finite momentum. 

The inclusion of both sources of dissipation provides a versatile parametric space in which to find all these phenomena, particularly polaritons with negative mass at selected detuning or momentum values. Depending on the model, one can consider $\gamma_{x}$, $\gamma_{c}$ and $\Omega_{\text{Im}}$ as independent parameters, e.g., if there are three environments one producing photon losses, another responsible of exciton losses, and a common environment for photons and exciton that gives rise to dissipative coupling. Otherwise, if one considers decays to a single bath, an input-output theory predicts that $\Omega_{\text{Im}}=\sqrt{\gamma_{x}\gamma_{x}}$~\cite{Bleu2024}. Other models have employed a similar approach~\cite{Wurdack2023}.

\begin{figure*}[!ht]
\begin{center}
\begin{tabular}{c}
\includegraphics[width=0.7\textwidth]{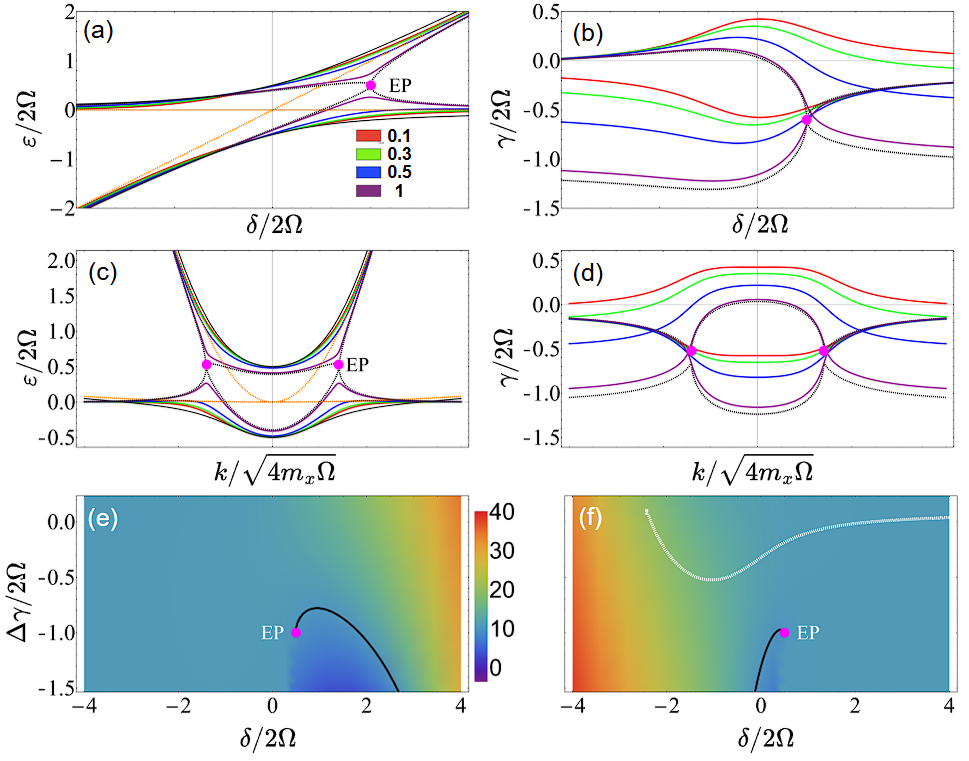}
\end{tabular}
\end{center}
\vspace{-20pt}
\caption{In panels (a)-(d) we show the same as Fig.~\ref{fig:1} (a)-(d), but for different values of the $\gamma_x/2\Omega=0.1,0.3,0.5$ and $1$, with $\gamma_{c}/2\Omega=0.1$ and $\Omega=\Omega_{\text{Im}}$. In panels (e) and (f), we show the same as in Fig.~\ref{fig:1} (e) and (f), where we have taken $\gamma_{x}=0.5(2\Omega)$. The BIC condition is plotted as a white curve.}
\label{fig:3} 
\end{figure*}

In Fig.~\ref{fig:3} we plot the results considering both $\Delta\gamma\neq0$ and $\Omega_{\text{Im}}\neq 0$. First, in Figs.~\ref{fig:3} (a) and (b), we observe that including all the effects shifts the level attraction to positive detuning. This is appreciated not only in the real and imaginary parts of the eigenvalues, but also in the position of the EP. The EPs now satisfy the complex degeneracy condition
\begin{gather}
(\delta_{\mathbf k}-i\Delta\gamma)^2+4(\Omega-i\Omega_{\mathrm{Im}})^2=0,
\end{gather}
leading to the equations 
\begin{gather}
\delta_{\mathbf{k}}^{\text{EP}}
=\pm\sqrt{\Delta\gamma^{2}-4(\Omega^{2}-\Omega_{\mathrm{Im}}^{2})},\\ \nonumber
\Delta\gamma^{\text{EP}}
=-\frac{4\Omega\Omega_{\mathrm{Im}}}{\delta_{\mathbf{k}}^{\text{EP}}},
\end{gather}
By substituting one into the other we obtain a fourth-degree equation for $\delta_{\textbf{k}}^{\text{EP}}$. When we solve this equation, we obtain a general condition for the existence of EPs that reads as:
\begin{gather} \label{eq:epg}
\delta_{\textbf{k}}^{\text{EP}}=\pm2\Omega_{\text{Im}},\,\,\,\,\Delta\gamma^{\text{EP}}=\mp2\Omega.
\end{gather}
These relations show that dissipative coupling shifts the EP away from zero detuning and modifies the balance between coherent and dissipative channels.

Second, in Figs.~\ref{fig:3} (c) and (d), we observe that it is easier to modify the curvature of the dispersion relations, so the area of negative mass in the parameter space [Fig.~\ref{fig:3} (e)] is larger for the upper polariton. Moreover, EPs appear in the dispersion relation as well, as for Eqs.~\ref{eq:epg}, their existence is also momentum-dependent. The change in curvature occurs at the EP. This means that, the presence of dissipative coupling in the decaying system induces the appearance of two EPs in the dispersion relation that strongly modifies the vicinity of the dispersion relation in parametric space, changing the effective mass.

The onset of zero and negative effective mass is also affected by dissipative coupling. In comparison to Figs.~\ref{fig:1} (e) and (f), in Figs.~\ref{fig:3} (e) and (f), we see that the dissipative coupling makes the negative mass region wider and overall increases the effective mass. 

Finally, to find BIC we again make zero $\gamma_{\sigma\mathbf{k}}$ and obtain
\begin{gather}
2\left(\gamma_x+\gamma_c\right)^2=\\ \nonumber
\sqrt{\left[\delta_{\textbf{k}}^{2}-\Delta\gamma^{2}+4(\Omega^{2}-\Omega_{\text{Im}}^{2})\right]^{2}+\left[2\delta_{\textbf{k}}\Delta\gamma+8\Omega\Omega_{\text{Im}}\right]^{2}}\\ \nonumber
-(\delta_{\textbf{k}}^{2}-\Delta\gamma^{2}+4(\Omega^{2}-\Omega_{\text{Im}}^{2})).
\end{gather} 
So, the quadratic equation determining the presence of BIC is
\begin{gather}\label{eq:BICt}
 \delta_{\mathbf{k}}^{2}
-\frac{2\Delta\gamma\,\Omega\Omega_{\mathrm{Im}}}{\gamma_{c}(\gamma_{c}-\Delta\gamma)}\,
\delta_{\mathbf{k}}\\ \nonumber
+\frac{(2\gamma_{c}-\Delta\gamma)^{2}}
      {\gamma_{c}(\gamma_{c}-\Delta\gamma)}
\left[\gamma_{c}(\gamma_{c}-\Delta\gamma)+(\Omega^{2}-\Omega_{\mathrm{Im}}^{2})\right] \\ \nonumber
-\frac{4\Omega^{2}\Omega_{\mathrm{Im}}^{2}}
       {\gamma_{c}(\gamma_{c}-\Delta\gamma)}=0 .
\end{gather}
This equation has two solutions in the parameter space; however, we only take the one with $\Omega_{\text{Im}}>0$. We plot this expression in Fig.~\ref{fig:3} (f) as a white curve. We observe that it does not seem to be correlated with the behavior of the effective mass. The presence of BIC is an effect only possible due to the combination of the dissipative coupling and the decay. 

In this regime, the imaginary part of the lower polariton branch may become negative, as shown in Figs.~\ref{fig:3} (b) and (d), indicating the onset of a gain instability: in the absence of sufficiently strong intrinsic losses, the dissipative coupling term injects energy into the polariton mode rather than extracting it. This behavior is different from a BIC, which would require the imaginary part of the eigenvalue to vanish exactly.

\begin{figure}[!ht]
\begin{center}
\begin{tabular}{c}
\includegraphics[width=1.0\columnwidth]{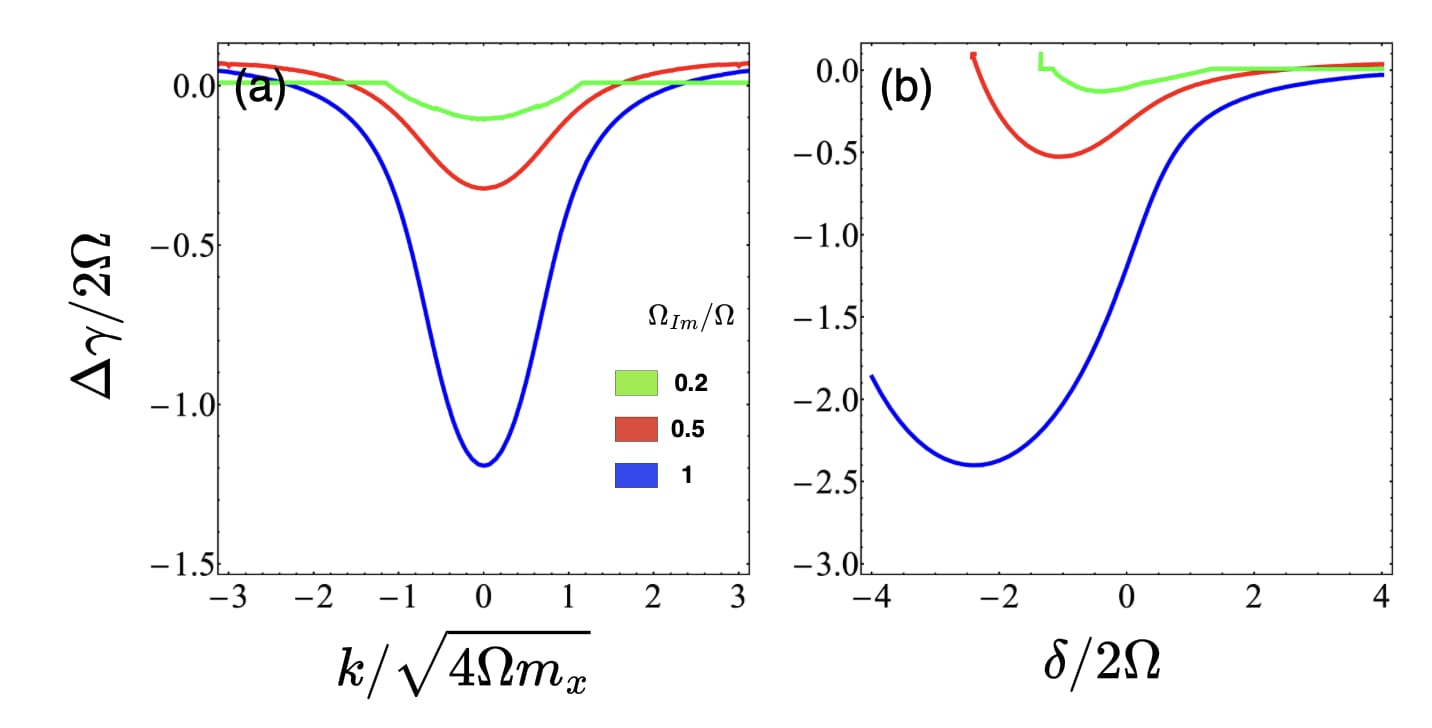}
\end{tabular}
\end{center}
\vspace{-20pt}
\caption{Parametric solution for BIC as a function of (a) detuning and (b) momentum for $\Omega_{\text{Im}}/\Omega=0.2$, $0.5$, and $1.0$ as indicated in the figure.}
\label{fig:4} 
\end{figure}

To identify the parameter regions where a true BIC can occur, we analyze Eq.~\ref{eq:BICt} as a function of detuning and momentum. The resulting structure is shown in Fig.~\ref{fig:4}. As shown in Fig.~\ref{fig:4}(a), the BIC condition disappears continuously when either $\Delta\gamma$ or $\Omega_{\mathrm{Im}}$ tends to zero, in agreement with the previous two cases. When both dissipation channels are present simultaneously, the condition defined by Eq.~\ref{eq:BICt} separates two qualitatively distinct regimes of the lower polariton branch. Increasing the dissipative coupling $\Omega_{\mathrm{Im}}$ shifts the BIC curve toward more negative values of $\Delta\gamma$ (i.e., $\gamma_{x}>\gamma_{c}$), which is the physically relevant regime for typical exciton–polariton systems. As a function of detuning, Fig.~\ref{fig:4}(b) shows that the BIC emerges predominantly at negative detuning, where the lower polariton becomes more photonic. In contrast, the upper branch does not admit a solution of the BIC condition and therefore cannot host a bound state in the continuum.

\section{Classical limit}
\label{sec:4}

In this section, we study the classical limit of the dissipative-coupling polariton Hamiltonian by taking the mean-field limit in terms of Glauber coherent states. In Ref.~\cite{Harder2021}, a Lagrangian approach is proposed to understand dissipative coupling physics; however, here we depart from the Hamiltonian expressed in the polariton basis, where we will explicitly have separated the real and imaginary contributions in Eqs.~\ref{eq:1} and~\ref{eq:2}. The generalized equations of motion for the expectation value of a non-Hermitian operator $\hat{A}$ in an arbitrary state $|\Psi\rangle$ are ($\hbar=1$)~\cite{Graefe2010}

\begin{gather}
\frac{d}{dt}\langle \hat{A}\rangle 
= i\langle[\hat{H},\hat{A}]\rangle
+\langle\{\hat{K},\hat{A}\}\rangle
+2\langle\hat{K}\rangle\langle\hat{A}\rangle .
\end{gather}
where $\{\dot,\dot\}$ is the anticommutator. Here, one has to consider that the norm of the initial state $n=\langle\Psi(t)|\Psi(t)\rangle$ is time-dependent and change with time as
$\dot{n}=-2\langle\psi(t)|\hat{K}|\psi(t)\rangle$~\cite{Graefe2010}. Particularly, we are interested in calculating the evolution of the annihilation (and creation) operators 
\begin{gather}
\frac{d}{dt}\langle\hat{L}_{\mathbf{k}}\rangle=-i\left\langle \frac{\partial \hat{H}}{\partial \hat{L}_{\mathbf{k}}^{\dagger}}\right\rangle-\left\langle\frac{\partial \hat{K}}{\partial \hat{L}_{\mathbf{k}}^{\dagger}}\right\rangle+2\left(\langle\hat{K}\rangle\langle\hat{L}_{\mathbf{k}}\rangle-\langle\hat{K}\hat{L}_{\mathbf{k}}\rangle\right),\nonumber\\  \nonumber
\frac{d}{dt}\langle\hat{L}_{\mathbf{k}^{\dagger}}\rangle=i\left\langle \frac{\partial \hat{H}}{\partial \hat{L}_{\mathbf{k}}}\right\rangle-\left\langle\frac{\partial \hat{K}}{\partial \hat{L}_{\mathbf{k}}}\right\rangle+2\left(\langle\hat{K}\rangle\langle\hat{L}_{\mathbf{k}}^{\dagger}\rangle-\langle\hat{K}\hat{L}_{\mathbf{k}}^{\dagger}\rangle\right),\\ \nonumber
\frac{d}{dt}\langle\hat{U}_{\mathbf{k}}\rangle=-i\left\langle \frac{\partial \hat{H}}{\partial \hat{U}_{\mathbf{k}}^{\dagger}}\right\rangle-\left\langle\frac{\partial \hat{K}}{\partial \hat{U}_{\mathbf{k}}^{\dagger}}\right\rangle
+2\left(\langle\hat{K}\rangle\langle\hat{U}_{\mathbf{k}}\rangle-\langle\hat{K}\hat{U}_{\mathbf{k}}\rangle
\right),\\ \nonumber
\frac{d}{dt}\langle\hat{U}_{\mathbf{k}}^{\dagger}\rangle=i\left\langle \frac{\partial \hat{H}}{\partial \hat{U}_{\mathbf{k}}}\right\rangle-\left\langle\frac{\partial \hat{K}}{\partial \hat{U}_{\mathbf{k}}}\right\rangle+
2\left(\langle\hat{K}\rangle\langle\hat{U}_{\mathbf{k}}^{\dagger}\rangle-\langle\hat{K}\hat{U}_{\mathbf{k}}^{\dagger}\rangle\right).
\end{gather}
We observe that the last two terms in the above equations contain the correlation between the time-changing operator of interest and the non-Hermitian part of the Hamiltonian. i.e., $\langle\hat{K}\rangle\langle\cdot\rangle-\langle\hat{K}\cdot\rangle$, which one expects to vanish in the mean-field approximation. Next, we employ Glauber coherent states $\hat{L}_{\mathbf{k}}|\ell_{\mathbf{k}}\rangle=\ell_{\mathbf{k}}|\ell_{\mathbf{k}}\rangle$ and $\hat{U}_{\mathbf{k}}|u_{\mathbf{k}}\rangle=u_{\mathbf{k}}|u_{\mathbf{k}}\rangle$ to obtain the corresponding classical equations of motion where $h_{\text{cl}}=\sum_{\mathbf{k}}\langle \ell_{\mathbf{k}}|\hat{H}_{\mathbf{k}}|\ell_{\mathbf{k}}\rangle$, and $k_{\text{cl}}=\sum_{\mathbf{k}}\langle \ell_{\mathbf{k}}|\hat{K}_{\mathbf{k}}|\ell_{\mathbf{k}}\rangle$ are the Weyl symbols of the $\hat{H}$ and $\hat{K}$ operators, respectively.  

The equations of motion become 
\begin{gather}
\dot{\ell_{\mathbf{k}}}=-i\frac{\partial h_{\text{cl}}}{\partial \ell_{\mathbf{k}}^{*}}-\frac{\partial k_{\text{cl}}}{\partial \ell_{\mathbf{k}}^{*}},\,\,\,\,
\frac{d\ell_{\mathbf{k}}^{*}}{dt}=i\frac{\partial h_{\text{cl}}}{\partial \ell_{\mathbf{k}}}-\frac{\partial k_{\text{cl}}}{\partial \ell_{\mathbf{k}}},\\
\dot{u}_{\mathbf{k}}=-i\frac{\partial h_{\text{cl}}}{\partial u_{\mathbf{k}}^{*}}-\frac{\partial k_{\text{cl}}}{\partial u_{\mathbf{k}}^{*}},\,\,\,\,
\dot{u}_{\mathbf{k}}^{*}=i\frac{\partial h_{\text{cl}}}{\partial u_{\mathbf{k}}}-\frac{\partial k_{\text{cl}}}{\partial u_{\mathbf{k}}}.
\end{gather}
However, it is convenient to write down the equations above in terms of canonical classical coordinates $(q_{\sigma\mathbf{k}},p_{\sigma\mathbf{k}})$ such that $\ell_\mathbf{k}=q_{\text{LP}\mathbf{k}}+ip_{\text{LP}\mathbf{k}}$ and $u_\mathbf{k}=Q_{\text{UP}\mathbf{k}}+iP_{\text{UP}\mathbf{k}}$. The corresponding classical equations of motion are 
\begin{gather}
\dot{q}_{\sigma\mathbf{k}}=\frac{\partial h_{\text{cl}}}{\partial p_{\sigma\mathbf{k}}}-\frac{\partial k_{\text{cl}}}{\partial q_{\sigma\mathbf{k}}},
\,\,\,\,
\dot{p}_{\sigma\mathbf{k}}=-\frac{\partial h_{\text{cl}}}{\partial q_{\sigma\mathbf{k}}}-\frac{\partial k_{\text{cl}}}{\partial p_{\sigma\mathbf{k}}},
\end{gather}
With our Hamiltonian in Eqs.~\ref{eq:1} we obtain
\begin{gather}\label{eqs:ham}
\dot{q}_{\sigma\mathbf{k}}=\epsilon_{\sigma\mathbf{k}}p_{\sigma\mathbf{k}}-\gamma_{\sigma\mathbf{k}}q_{\sigma\mathbf{k}},
\\ \nonumber
\dot{p}_{\sigma\mathbf{k}}=-\epsilon_{\sigma\mathbf{k}}q_{\sigma\mathbf{k}}-\gamma_{\sigma\mathbf{k}}p_{\sigma\mathbf{k}}.
\end{gather}
If $\gamma_{\sigma\mathbf{k}}>0$, these equations correspond to a damped harmonic oscillator
\begin{gather}
\ddot{q}_{\sigma\mathbf{k}}+2\gamma_{\sigma\mathbf{k}}\dot{q}_{\mathbf{k}}+\omega_{\sigma\mathbf{k}}^{2} q_{\sigma\mathbf{k}}=0,
\end{gather}
with frequencies 
\begin{gather}
\omega_{\sigma\mathbf{k}}^{2}=||E_{\sigma\mathbf{k}}||^{2}=\epsilon_{\sigma\mathbf{k}}^{2}+\gamma_{\sigma\mathbf{k}}^{2}=\\ \nonumber
\frac{1}{4}\left[\left(\varepsilon_{x\textbf{k}}+\varepsilon_{c\textbf{k}}\right)^2+\left(\gamma_x+\gamma_c\right)^2\right.
    \left.\right.
    \\ \nonumber
    \pm 2\left(\varepsilon_{x\textbf{k}}+\varepsilon_{c\textbf{k}}\right)\mathcal{F}_{\mathbf{k}}^{+}
     \left.\pm 2 \text{sgn}(\mathcal{J}_{\mathbf{k}})\left(\gamma_x+\gamma_c\right)\mathcal{G}_{\mathbf{k}}\mathcal{F}_{\mathbf{k}}^{-}\right].
\end{gather}

Note that the frequency $\omega_{\sigma\mathbf{k}}$ arises directly from the second-order equation obtained from the non-Hermitian first-order dynamics with eigenvalue  $E_{\sigma\mathbf{k}}$. Hence, in contrast to the textbook damped oscillator, where one postulates a second-order equation
$\ddot{x}+2\gamma\dot{x}+\omega_{0}^{2}x=0$ with $\gamma$, $\omega_{0}$, $\sqrt{\omega_{0}^{2}-\gamma^{2}}$ the decay, oscillator frequency, and oscillation frequency, respectively, here the quantity $\omega_{\sigma\mathbf{k}}$ is the modulus $|E_{\sigma\mathbf{k}}|$ and does \emph{not} represent the physical oscillation frequency, which is given instead by the real part $\varepsilon_{\sigma\mathbf{k}}$.

For the damping conditions, there is subcritical behavior, with $\omega_{\sigma\mathbf{k}}>\gamma_{\sigma\mathbf{k}}$. Instead, in the regimes where $\gamma_{\sigma\mathbf{k}}<0$  a repulsor emerges, as discussed in previous sections. 

The solution for the quadratures in each polariton branch is
\begin{gather} \label{eq:q}
q_{\sigma\mathbf{k}}(t)=q_{\sigma\mathbf{k}}(0)e^{-\gamma_{\sigma\mathbf{k}}t}\cos(\omega_{\sigma\mathbf{k}}t+\Phi_{\sigma}),
\end{gather}
where $\Phi_{\sigma}$ is an arbitrary phase. Notice that the quadrature is, on average, proportional to the square root of the polariton density $\sqrt{n_{\sigma\mathbf{k}}}$. First, we can look for fixed points in the dynamics. They come from making zero Eqs.~\ref{eqs:ham}, which lead to either $q_{\mathbf{k}}=p_{\mathbf{k}}=0$, or the condition $\gamma_{\sigma\mathbf{k}}^{2}+\epsilon_{\sigma\mathbf{k}}^{2}=0$, which cannot be held, given that both quantities are real.

\subsection{Temporal evolution of the quadrature}

\begin{figure*}[!ht]
\begin{center}
\begin{tabular}{c}
\includegraphics[width=0.7\textwidth]{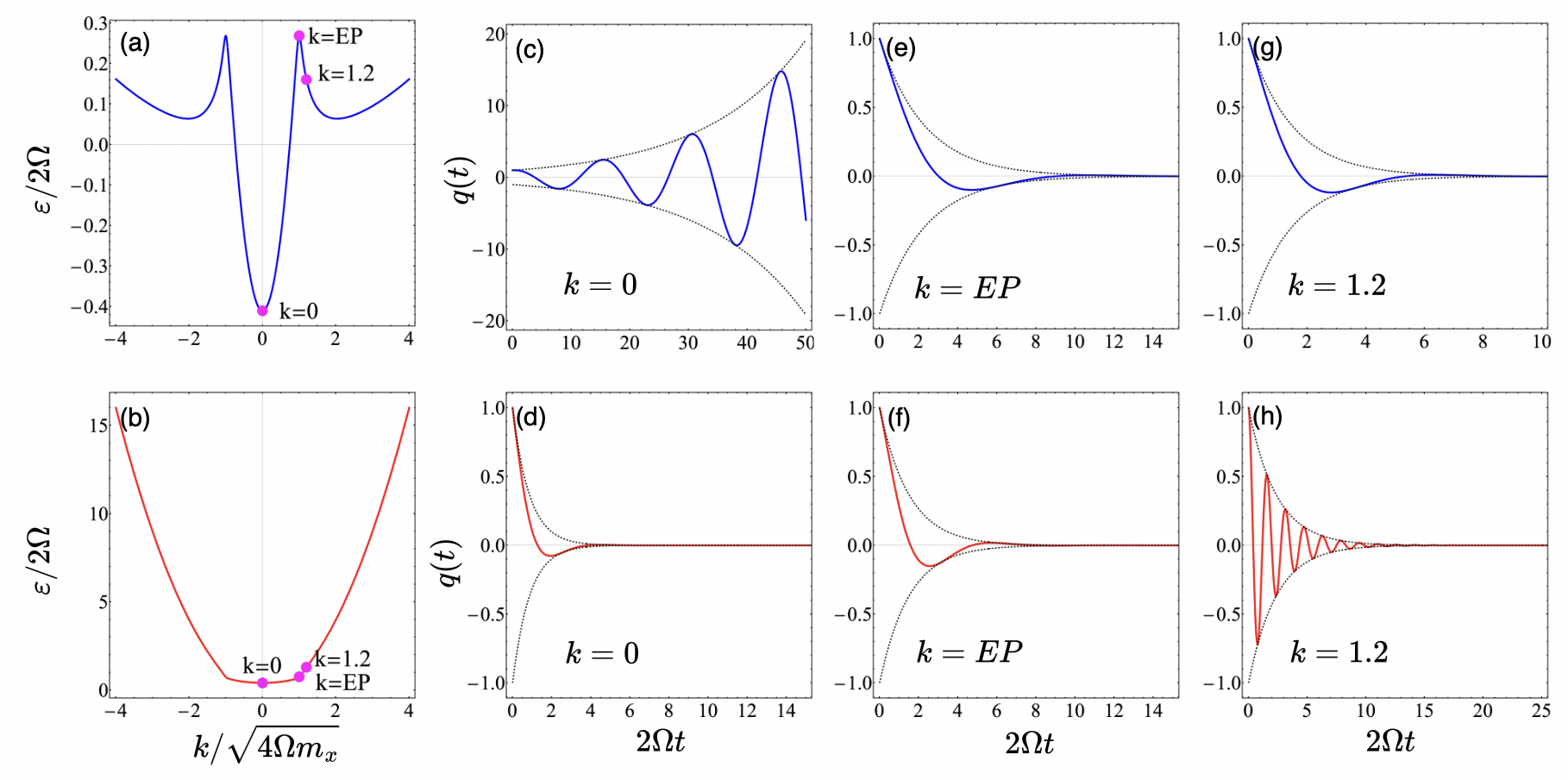}
\end{tabular}
\end{center}
\vspace{-20pt}
\caption{Temporal evolution of the classical polariton quadrature $q_{\sigma\mathbf{k}}(t)$ for the (top row) upper and (bottom row) lower polaritons at $\Omega_{\text{Im}}=\Omega$ and different values of relative dissipation $\Delta\gamma$ and $k$ as indicated in (a).}
\label{fig:5} 
\end{figure*}

In Fig.~\ref{fig:5} we explore the time evolution of $q_{\sigma\mathbf{k}}$ for selected values of momentum and a representative value of the dissipative coupling $\Omega_{\text{Im}}=\Omega$, for $\gamma_{c}/2\Omega=0.1$, $\gamma_{x}/2\Omega=0.1$, and at $\delta=0.0$. In Fig.~\ref{fig:5} (a) and (b), we show the dispersion relation of the lower and upper polaritons, respectively, where we have selected three representative values of momentum's norm: $k=0$, the momentum of the EP, and $k=2$. For the lower polariton, there is a change in behavior between Figs.~\ref{fig:5} (c) and (e), as in the first case, the imaginary part becomes negative, so the quadrature increases in time, but in Fig.~\ref{fig:5} (e) and (g), one finds the standard damped oscillator with subcritical behavior, with the quadrature rapidly vanishing. This happens once one reaches the change in curvature of the dispersion relation, marked by the EP. Instead, the upper polariton always has a dissipative behavior, as seen from Figs.~\ref{fig:5} (d), (f), and (g). The main difference when we cross the EP is the manifestation of quick oscillations before the upper polariton dissipates. 

Whether one has an attractor or repulsor depends on the sign of $\gamma_{\sigma\mathbf{k}}$. However, by taking the derivatives of Eq.~\ref{eq:q}, where we fix $\Phi_{\sigma}=0$ and $q_{\sigma\mathbf{k}}(0)=q_{0}$ without loss of generality, one can identify the effects of the  on the time evolution of the quadrature. The second derivative reads
\begin{gather}\label{eq:second}
\nabla_{\mathbf{k}}^{2}q_{\sigma\mathbf{k}}(t)=-q_{0}e^{-\gamma_{\sigma\mathbf{k}}t}\left\{t^{2}\mathcal{Q}_{\mathbf{k}}(t)+t\mathcal{P}_{\mathbf{k}}(t)+
\right.
\\ \nonumber
\left.
t\frac{\sin(\omega_{\sigma\mathbf{k}}t)}{\omega_{\sigma\mathbf{k}}}\left[\nabla^{2}_{\mathbf{k}}\gamma_{\sigma\mathbf{k}}\left(\omega_{\sigma\mathbf{k}}\cot(\omega_{\sigma\mathbf{k}}t)+\gamma_{\sigma\mathbf{k}}\right)+\epsilon_{\sigma\mathbf{k}}\nabla^{2}_{\mathbf{k}}\epsilon_{\sigma\mathbf{k}}
\right]
\right\},
\end{gather}
where $\mathcal{Q}_{\mathbf{k}}(t)$ and $\mathcal{P}_{\mathbf{k}}(t)$ are oscillating functions that are defined in App.~\ref{app:2}. We observe two main contributions proportional to $\sim t$ and $\sim t^{2}$. The terms comprising the information of the diffusive mass, i.e., those depending on the second derivative of the eigenvalues, are linear terms in time, so they are relevant at short times. Therefore, as a function of momentum, the negative mass is reflected in the short time scale of the polaritons.

\section{Conclusions}
\label{sec:5}

We have performed a parametric analysis of the impact of dissipative effects on the excitations of an exciton-polariton system. By treating the dissipative coupling and the exciton and photon decay rates independently, we have derived exact relations for the upper and lower polariton branches and characterized the effect of each parameter, inducing a non-Hermitian effect.

In the absence of dissipative coupling, the relative dissipation between light and matter becomes the relevant parameter, inducing level attraction and a change in curvature of the dispersion relation, thus a change in sign of the effective mass at zero momentum, where the condition $\Delta\gamma^{\text{EP}}=\pm2\Omega$ establishes a limiting value, at which the exceptional points (EPs) appear. They are a result of the level attraction between the lower and upper polaritons at maximum hybridization. On the contrary, in the absence of external dissipation channels, dissipative coupling does not induce EPs but creates level attraction due to an energy shift produced by the light-matter detuning. It also produces a change in the curvature of the polariton dispersion at higher values of momentum, something that has been experimentally observed before~\cite{Dhara2018,Wurdack2023}. We show that in the presence of dissipative coupling two EPs appear inducing a deformation in the parameter space resulting in a stronger deformation of the polariton dispersion. The combination of both non-Hermitian sources allows for shifting the position of the EPs as a function of detuning, given the condition $\delta_{\mathbf{k}}^{EP}=2\Omega_{\text{Im}}$. Hence, this allows for shifting the position of the level attraction and simultaneously, manipulating the curvature of the mass parameters. On the other hand, only in the presence of both dissipative coupling and decay channels do one find non-decaying polariton states, i.e., BICs. They set the condition for the vanishing of the imaginary part of the polariton branches. This means it can become negative, signaling an unstable behavior in the polaritons. 

Finally, we have explored the classical limit from the Hamiltonian approach using coherent states. This allows us to observe the quadrature's behavior over time. This confirms the quantum behavior, as in general we get an attractor or repulsor depending on the sign of the imaginary part of the polariton branches. This behavior depends on momentum and the system's non-Hermitian parameters. In particular, the negative mass effects associated with a change in the sign of curvature are noticeable only at short times, as one observes subcritical damped-oscillator behavior later on.  

Therefore, our analysis has unveiled the parameter conditions for the onset of non-Hermitian phenomena, such as EPs and BICs, in polariton systems, as well as the anomalous behavior of polariton dispersion relations. We expect this to contribute establishing a roadmap for understanding non-Hermitian effects in many-body polariton states. 

\section*{Acknowledgements} 
The authors acknowledge financial support from CONAHCYT/Secihti No. CBF2023-2024-1765. A. J. V. C. acknowledges support from the Graduate Program scholarship from CONAHCYT/Secihti. M. A. B. M. acknowledges financial support from the PIPAIR 2024 project from the DAI UAM, and the Marcos Moshinsky Fellowship. A.C.G acknowledges support from PIIF25 and UNAM DGAPA PAPIIT Grant No. IA101325


\appendix
\unskip
\begin{widetext}
\section{Derivatives of the real and imaginary parts of polariton branches}
\label{app:1}

The first derivative of the real and imaginary of the eigenenergies in Eq.~\ref{eq:d1} from the main text are given by
\begin{gather}
    \mathbf{k}\cdot\nabla_{\mathbf{k}}\varepsilon_{\sigma\textbf{k}}=
    \frac{1}{2}\left(\frac{\textbf{k}^{2}}{m_c}+\frac{\textbf{k}^{2}}{m_x}\right) \pm\frac{1}{2}\left(\frac{\textbf{k}^{2}}{m_c}-\frac{\textbf{k}^{2}}{m_x}\right)\frac{1}{2\mathcal{F}_{\mathbf{k}}^{+}} 
    \left\{\frac{1}{\mathcal{G}_{\mathbf{k}}}\left[\mathcal{H}_{\mathbf{k}}\delta_{\textbf{k}}+\mathcal{J}_{\mathbf{k}}\Delta\gamma\right]+\delta_{\textbf{k}}\right\},
    \end{gather}
and
\begin{gather}
    \mathbf{k}\cdot\nabla_{\mathbf{k}}\gamma_{\sigma\textbf{k}}=\pm \left(\frac{\textbf{k}^{2}}{m_c}-\frac{\textbf{k}^{2}}{m_x}\right) \frac{\text{sign}\left[\mathcal{J}_{\mathbf{k}}\right]}{4\mathcal{F}_{\mathbf{k}}^{-}}
    \times\left\{\frac{1}{\mathcal{G}_{\mathbf{k}}}\left[\mathcal{H}_{\mathbf{k}}\delta_{\textbf{k}}+\mathcal{J}_{\mathbf{k}}\Delta\gamma\right]-\delta_{\textbf{k}}\right\}.
\end{gather}
Likewise, the second derivative of the real and imaginary parts of the eigenenergies in Eq.~\ref{eq:m2} are
\begin{gather}
\nabla_{\textbf{k}}^{2}\varepsilon_{\sigma\textbf{k}}=\frac{1}{2}\left(\left(\frac{1}{m_c}+\frac{1}{m_x}\right)\right. 
    \left.\mp\frac{1}{2}\frac{1}{\mathcal{F}_{\mathbf{k}}^{+2}} \right. 
    \left[\frac{1}{\mathcal{G}_{\mathbf{k}}}\left(\mathcal{H}_{\mathbf{k}}
\delta_{\textbf{k}}+\mathcal{J}\Delta\gamma\right)+\delta_{\textbf{k}}\frac{}{}\right]^{2}\left(\frac{\textbf{k}}{m_c}-\frac{\textbf{k}}{m_x}\right)^{2} \\ \nonumber
    \pm \frac{1}{\mathcal{F}^{+}_{\textbf{k}}}
     \left\{-\frac{1}{2}\frac{1}{\mathcal{G}_{\textbf{k}}^{3}}\right. \left(\mathcal{H}_{\textbf{k}}
    \delta_{\textbf{k}}+\mathcal{J}_{\textbf{k}}\Delta\gamma\right)^{2}\left(\frac{\textbf{k}}{m_c}-\frac{\textbf{k}}{m_x}\right)^{2} \\ \nonumber
    + \frac{1}{\mathcal{G}_{\textbf{k}}} \left[\left(2\delta_{\textbf{k}}^{2}
    +\mathcal{H}_{\textbf{k}}
    +\mathcal{H}_{\textbf{k}}\left(\frac{1}{m_c}-\frac{1}{m_x}\right)\delta_{\textbf{k}} 
    2\Delta\gamma^{2}\right)\left(\frac{\textbf{k}}{m_c}-\frac{\textbf{k}}{m_x}\right)^{2}+\mathcal{J}_{\textbf{k}}\Delta\gamma\left(\frac{1}{m_c}-\frac{1}{m_x}\right)\right] \\ \nonumber
   \left.\left.+\left(\frac{\textbf{k}}{m_c}-\frac{\textbf{k}}{m_x}\right)^{2}+\delta_{\textbf{k}}\left(\frac{1}{m_c}-\frac{1}{m_x}\right)\right\}\right)
\end{gather}

\begin{gather}
    \nabla_{\textbf{k}}^{2}\gamma_{\sigma\textbf{k}}=\frac{1}{4}\text{sign}\left[\mathcal{J}_{\textbf{k}}\right] 
    \left(\mp\frac{1}{2}\frac{1}{(\mathcal{F}^{-}_{\textbf{k}})^{3}} \right. 
    \left[\frac{1}{\mathcal{G}_{\textbf{k}}}\left(\mathcal{H}_{\textbf{k}}
    \delta_{\textbf{k}}+\mathcal{J}_{\textbf{k}}\Delta\gamma\right)-\delta_{\textbf{k}}\right]^{2}\left(\frac{\textbf{k}}{m_c}-\frac{\textbf{k}}{m_x}\frac{}{}\right)^{2} \\ \nonumber
    \pm \frac{1}{\mathcal{F}^{-}_{\textbf{k}}}
     \left\{-\frac{1}{2}\frac{1}{\mathcal{G}_{\textbf{k}}^{3}}\right. \left(\mathcal{H}_{\textbf{k}}\delta_{\textbf{k}}+\mathcal{J}_{\textbf{k}}\Delta\gamma\right)^{2}\left(\frac{\textbf{k}}{m_c}-\frac{\textbf{k}}{m_x}\right)^{2} 
   \\ \nonumber
     + \frac{1}{\mathcal{G}_{\textbf{k}}} \left[
     \left(2\delta_{\textbf{k}}^{2} +\mathcal{H}_{\textbf{k}}+\mathcal{H}_{\textbf{k}}\left(\frac{1}{m_c}-\frac{1}{m_x}\right)\delta_{\textbf{k}} 
    2\Delta\gamma^{2}\right)\left(\frac{\textbf{k}}{m_c}-\frac{\textbf{k}}{m_x}\right)^{2} +\mathcal{J}_{\textbf{k}}\Delta\gamma\left(\frac{1}{m_c}-\frac{1}{m_x}\right)\right] 
    \\ \nonumber 
   \left.\left.-\left(\frac{\textbf{k}}{m_c}-\frac{\textbf{k}}{m_x}\right)^{2}-\delta_{\textbf{k}}\left(\frac{1}{m_c}-\frac{1}{m_x}\right)\right\}\right)
\end{gather}

\section{Derivative of the quadrature as a function of momentum}
\label{app:2}

The first derivative reads
\begin{gather}
\nabla_{\mathbf{k}}q_{\sigma\mathbf{k}}(t)=-tq_{0}e^{-\gamma_{\sigma\mathbf{k}}t}
\left[
(\nabla_{\mathbf{k}}\gamma_{\sigma\mathbf{k}})
\cos(\omega_{\sigma\mathbf{k}}t)
+(\nabla_{\mathbf{k}}\omega_{\sigma\mathbf{k}})\sin(\omega_{\sigma\mathbf{k}}t)
\right],
\end{gather}
given that
\begin{gather}\label{eq:nabom}
\nabla_{\mathbf{k}}\omega_{\sigma\mathbf{k}}=\frac{1}{\sqrt{\epsilon_{\sigma\mathbf{k}}^{2}+\gamma_{\sigma\mathbf{k}}^{2}}}\left(\epsilon_{\mathbf{k}}\nabla_{\mathbf{k}}\epsilon_{\mathbf{k}}+\gamma_{\mathbf{k}}\nabla_{\mathbf{k}}\gamma_{\mathbf{k}}\right)=\frac{\epsilon_{\mathbf{k}}}{\omega_{\sigma\mathbf{k}}}\nabla_{\mathbf{k}}\epsilon_{\mathbf{k}}+\frac{\gamma_{\mathbf{k}}}{\omega_{\sigma\mathbf{k}}}\nabla_{\mathbf{k}}\gamma_{\mathbf{k}}
\end{gather}
the above expression can be rearranged as 
\begin{gather}
\nabla_{\mathbf{k}}q_{\sigma\mathbf{k}}(t)=-tq_{0}e^{-\gamma_{\sigma\mathbf{k}}t}
\left[
\nabla_{\mathbf{k}}\gamma_{\sigma\mathbf{k}}\left[\cos(\omega_{\sigma\mathbf{k}}t)+
\frac{\gamma_{\sigma\mathbf{k}}}{\omega_{\sigma\mathbf{k}}}\,\sin(\omega_{\mathbf{k}\sigma}t)
\right] \right.
\left. +\nabla_{\mathbf{k}}\epsilon_{\sigma\mathbf{k}}\frac{\epsilon_{\sigma\mathbf{k}}}{\omega_{\sigma\mathbf{k}}}\,
\sin(\omega_{\sigma\mathbf{k}}t)
\right].
\end{gather}
Meanwhile, the second derivative is
\begin{gather}
\nabla_{\mathbf{k}}^{2}q_{\sigma\mathbf{k}}(t)=-q_{0}e^{-\gamma_{\sigma\mathbf{k}}t}t^{2}\left\{
\left[-(\nabla_{\mathbf{k}}\gamma_{\sigma\mathbf{k}})^{2}(\cos(\omega_{\sigma\mathbf{k}})+\frac{\gamma_{\sigma\mathbf{k}}}{\omega_{\sigma\mathbf{k}}}\sin(\omega_{\sigma\mathbf{k}}))
-\nabla_{\mathbf{k}}\gamma_{\sigma\mathbf{k}}\cdot \nabla_{\mathbf{k}}\epsilon_{\sigma\mathbf{k}} \frac{\epsilon_{\sigma\mathbf{k}}}{\omega_{\sigma\mathbf{k}}}
\sin(\omega_{\sigma\mathbf{k}})
\right]\right.
\nonumber \\ 
\left.
\nabla_{\mathbf{k}}\omega_{\sigma\mathbf{k}}\cdot\left[\frac{}{}-\nabla_{\mathbf{k}}\gamma_{\sigma\mathbf{k}}\right[\sin(\omega_{\sigma\mathbf{k}}t)-\frac{\gamma_{\sigma\mathbf{k}}}{\omega_{\sigma\mathbf{k}}}\cos(\omega_{\sigma\mathbf{k}}t)\left]+\nabla_{\mathbf{k}}\epsilon_{\sigma\mathbf{k}}\frac{\epsilon_{\sigma\mathbf{k}}}{\omega_{\sigma\mathbf{k}}}\cos(\omega_{\sigma\mathbf{k}}t)
\right]
\right\}\\ \nonumber
-q_{0}e^{-\gamma_{\sigma\mathbf{k}}t}t\left\{\nabla^{2}_{\mathbf{k}}\gamma_{\sigma\mathbf{k}}\left(
\cos(\omega_{\sigma\mathbf{k}}t)+\frac{\gamma_{\sigma\mathbf{k}}}{\omega_{\sigma\mathbf{k}}}\sin(\omega_{\sigma\mathbf{k}}t)
\right)+ \nabla_{\mathbf{k}}^{2}\epsilon_{\sigma\mathbf{k}}\frac{\epsilon_{\sigma\mathbf{k}}}{\omega_{\sigma\mathbf{k}}}\sin(\omega_{\sigma\mathbf{k}}t)\right.\\ \nonumber
\left.
\left[(\nabla_{\mathbf{k}}\epsilon_{\sigma\mathbf{k}})^{2}
+ (\nabla_{\mathbf{k}}\gamma_{\sigma\mathbf{k}})^{2}
\right]
\frac{\sin(\omega_{\sigma\mathbf{k}}t)}{\omega_{\sigma\mathbf{k}}}
-\nabla_{\mathbf{k}}\omega_{\sigma\mathbf{k}}\cdot\left[\frac{\epsilon_{\sigma\mathbf{k}}}{\omega_{\sigma\mathbf{k}}}\nabla_{\mathbf{k}}\epsilon_{\sigma\mathbf{k}}
+\gamma_{\sigma\mathbf{k}}\nabla_{\mathbf{k}}\gamma_{\sigma\mathbf{k}}
\right]\frac{\sin(\omega_{\sigma\mathbf{k}}t)}{\omega_{\sigma\mathbf{k}}}
\right\}.
\end{gather}
We notice there are terms of linear and quadratic order in $t$. Next, we substitute the derivative of $\omega_{\sigma\mathbf{k}}$ in Eq.~\ref{eq:nabom} to obtain Eq.~\ref{eq:second} from the main text. It reads, 
\begin{gather}
\nabla_{\mathbf{k}}^{2}q_{\sigma\mathbf{k}}(t)=-q_{0}e^{-\gamma_{\sigma\mathbf{k}}t}\left\{t^{2}\mathcal{Q}_{\mathbf{k}}(t)+t\mathcal{P}_{\mathbf{k}}(t)+
\right.
\\ \nonumber
\left.
t\frac{\sin(\omega_{\sigma\mathbf{k}}t)}{\omega_{\sigma\mathbf{k}}}\left[\nabla^{2}_{\mathbf{k}}\gamma_{\sigma\mathbf{k}}\left(\omega_{\sigma\mathbf{k}}\cot(\omega_{\sigma\mathbf{k}}t)+\gamma_{\sigma\mathbf{k}}\right)+\epsilon_{\sigma\mathbf{k}}\nabla_{\mathbf{k}}\epsilon_{\sigma\mathbf{k}}
\right]
\right\}
\end{gather}
where
\begin{gather}
\mathcal{Q}_{\mathbf{k}}(t)=
\left\{
\left(\nabla_{\mathbf{k}}\epsilon_{\sigma\mathbf{k}}\right)^{2}\frac{\epsilon_{\mathbf{k}}^{2}}{\omega_{\mathbf{k}}^{2}}\cos(\omega_{\sigma\mathbf{k}}t)+
\right.
\\ \nonumber
\left.
\left(\nabla_{\mathbf{k}}\gamma_{\sigma\mathbf{k}}\right)^{2}\left[\left(\frac{\gamma_{\mathbf{k}}^{2}}{\omega_{\mathbf{k}}^{2}}-1\right)\omega_{\sigma\mathbf{k}}\cot(\omega_{\mathbf{k}}t)-2\gamma_{\sigma\mathbf{k}}\right]\frac{\sin(\omega_{\sigma\mathbf{k}}t)}{\omega_{\sigma\mathbf{k}}}+ 
\right.
\\ \nonumber
\left.
\nabla_{\mathbf{k}}\epsilon_{\sigma\mathbf{k}}\cdot\nabla_{\mathbf{k}}\gamma_{\sigma\mathbf{k}}\left(\frac{\gamma_{\mathbf{k}}}{\omega_{\sigma\mathbf{k}}}\cot(\omega_{\sigma\mathbf{k}}t)-1\right)\frac{2\epsilon_{\sigma\mathbf{k}}}{\omega_{\sigma\mathbf{k}}}\sin(\omega_{\sigma\mathbf{k}}t)\right\},\\
\mathcal{P}_{\mathbf{k}}(t)=\frac{\sin(\omega_{\sigma\mathbf{k}}t)}{\omega_{\sigma\mathbf{k}}}
\left\{
\left(\nabla_{\mathbf{k}}\gamma_{\sigma\mathbf{k}}\right)^{2}\left(1-\frac{\gamma_{\sigma\mathbf{k}}^{2}}{\omega_{\sigma\mathbf{k}}}\right)+
\left(\nabla_{\mathbf{k}}\epsilon_{\sigma\mathbf{k}}\right)^{2}\left(1-\frac{\epsilon_{\sigma\mathbf{k}}^{2}}{\omega_{\sigma\mathbf{k}}}\right)\right.
\\ \nonumber
\left.
-2\frac{\epsilon_{\sigma\mathbf{k}}\omega_{\gamma\mathbf{k}}}{\omega_{\sigma\mathbf{k}}}\nabla_{\mathbf{k}}\epsilon_{\sigma\mathbf{k}}\cdot\nabla_{\mathbf{k}}\gamma_{\sigma\mathbf{k}}
\right\}.
\end{gather}
\end{widetext}


\bibliography{references}

\end{document}